\documentclass{amsart}
\usepackage{color,graphicx,subfigure}
\usepackage{algorithm}
\usepackage{algpseudocode}
\algrenewcommand{\algorithmiccomment}[1]{\hfill(#1)}

\date{\today}
\title[Accelerated density matrix expansions]{Accelerated density matrix expansions for Born--Oppenheimer molecular dynamics}

\author[E. H. Rubensson]{Emanuel H. Rubensson}
\email[E. H. Rubensson]{emanuel.rubensson@it.uu.se}
\address[E. H. Rubensson]{Division of Scientific Computing, Department of Information Technology, Uppsala University, Box 337, SE-751 05 Uppsala, Sweden}
\author[A. M. N. Niklasson]{Anders M. N. Niklasson}
\email[A. M. N. Niklasson]{amn@lanl.gov}
\address[A. M. N. Niklasson]{Theoretical division, Los Alamos National Laboratory, Los Alamos, New Mexico 87545, USA}

\begin{document}
\begin{abstract}
An accelerated polynomial expansion scheme to construct the density
matrix in quantum mechanical molecular dynamics simulations is
proposed. The scheme is based on recursive density matrix expansions,
e.g. [Phys. Rev. B.  66 (2002), p. 155115], which are accelerated by a
scale-and-fold technique [J. Chem. Theory Comput. 7 (2011), p. 1233].
The acceleration scheme requires interior eigenvalue estimates, which
may be expensive and cumbersome to come by.  Here we show how such
eigenvalue estimates can be extracted from the recursive expansion by
a simple and robust procedure at a negligible computational cost.  Our
method is illustrated with density functional tight-binding
Born--Oppenheimer molecular dynamics simulations, where the
computational effort is dominated by the density matrix construction.
In our analysis we identify two different phases of the recursive
polynomial expansion, the conditioning and purification phases, and we
show that the acceleration represents an improvement of the
conditioning phase, which typically gives a significant reduction of
the computational cost.
\end{abstract}

\maketitle

\section{Introduction} 

With the fast growth of computational processing power, atomistic
simulations based on quantum mechanical calculations of the electronic structure 
have become a powerful approach to the study of a broad range of problems
in materials science, chemistry, and biology
\cite{BKirchner12,DMarx00,MTuckerman02}.  Nevertheless, the
computational cost associated with electronic
structure calculations normally limits applications to
fairly small systems. In particular, the cubic, $O(N^3)$, scaling of
the computational cost as a function of the number of atoms, $N$, for
the regular solution of the quantum mechanical 
eigenvalue problem is considered to be a most limiting
factor. A number of different electronic structure technologies have
therefore been developed that circumvent this bottleneck with a
computational effort that scales only linearly with the system size
\cite{DBowler12,SGoedecker99}.  The reduction in the 
cost is typically achieved by utilizing sparse matrix algebra in an
iterative construction of the density matrix, which avoids the full
regular solution of the quantum mechanical eigenvalue problem.  The
matrix sparsity arise in localized atomic basis set representations due to
the short-range character of the electronic wavefunctions for
non-metallic materials \cite{MBenzi13,WKohn59,WKohn96}. With linear scaling
techniques it is today possible to perform electronic structure
calculations involving millions of atoms
\cite{DBowler10,JVandeVondele12}.

An ideal linear scaling method should combine several important
properties: {\em i)} it should be well tailored for high-performance
on heterogeneous multi-core architectures, {\em ii)} require little
intermediate memory storage, {\em iii)} allow well controlled
numerical accuracy, and {\em iv)} have a low computational
prefactor. Among several linear scaling techniques
\cite{multilevel-bchb, Birgin2013, dmm-c, dmm-d, pur-h, KimJung2011, Kim1995, dmm-lnv,
  pur-m, lo-odgm}, the recursive second-order spectral projection
method (SP2) \cite{pur-n}, which sometimes also is referred to as the
trace-correcting purification algorithm, represents a surprisingly
simple and efficient choice \cite{EHRubensson11}.  The SP2 algorithm
allows for control of errors arising from an approximate sparse matrix
algebra \cite{ANiklasson03, mixedNormTrunc, accPuri} and it has a low
computational prefactor. Even in the dense (non-sparse) limit the SP2
algorithm outperforms regular lapack diagonalization techniques on
multi-core platforms, both with respect to speed and accuracy
\cite{MCawkwell12GPU}. The computational cost of the SP2 algorithm is
dominated by a sequence of matrix-matrix multiplications; typically
between 20 to 40 multiplications are needed.  Highly optimized linear
algebra subroutines can perform dense matrix-matrix multiplications
with close to peak performance on multi-core platforms.  Thus, even
without utilizing linear scaling sparse matrix algebra, the SP2
algorithm has excellent performance and is well suited for modern
computational architectures \cite{MCawkwell12GPU}.

Recently Rubensson proposed an acceleration technique that can further
boost the performance of the SP2 algorithm, as well as other recursive
polynomial expansion schemes \cite{non-monotonic}. The acceleration is
based on the idea to allow for nonmonotonicity in the interval of
interest, which gives additional flexibility in the choice of
recursive expansion polynomials.  By the use of a scale and fold
technique, i.e. by shifting and re-scaling the SP2 projection polynomials, 
the number of matrix-matrix multiplications required by the
SP2 algorithm is significantly reduced.  This boosting
technique requires estimates of interior eigenvalues that may be
expensive and cumbersome to come by. In this article we show how such
eigenvalue estimates can be extracted by a simple procedure from the
recursive polynomial expansion at negligible computational cost.

Our scheme is particularly useful in molecular dynamics simulations.
A molecular dynamics simulation often involves hundreds of thousands of
time steps and therefore requires a low computational cost for the
calculation of each new configuration. This is provided by the
accelerated SP2 algorithm where eigenvalue estimates that are
necessary for the acceleration can be extracted from previous time steps
at practically no extra cost.  The scheme is illustrated with density
functional tight-binding molecular dynamics simulations, where the
computational effort is dominated by the construction of the
electronic density matrix through the recursive polynomial expansion.
All the simulations are performed using dense matrix algebra in the
computational framework of extended Lagrangian Born--Oppenheimer
molecular dynamics \cite{ANiklasson08,PSteneteg10}, which provides
accurate and stable molecular trajectories with long-term energy
conservation.

The article is outlined as follows: first we describe the problem and
the recursive SP2 expansion of the density matrix that is
used iteratively in the self-consistent calculation of the electronic
ground state.  Thereafter the acceleration technique is presented and
we show how the required interior eigenvalues can be estimated from
the recursion. The article goes on with a presentation of the extended
Lagrangian formulation of Born--Oppenheimer molecular dynamics and we
illustrate the behavior of our scheme in some quantum mechanical
molecular dynamics simulations before we end with a discussion and
some concluding remarks.

\section{The density matrix} \label{sec:density_matrix}
\begin{figure}
  \begin{center}
    \includegraphics[width=0.95\textwidth]{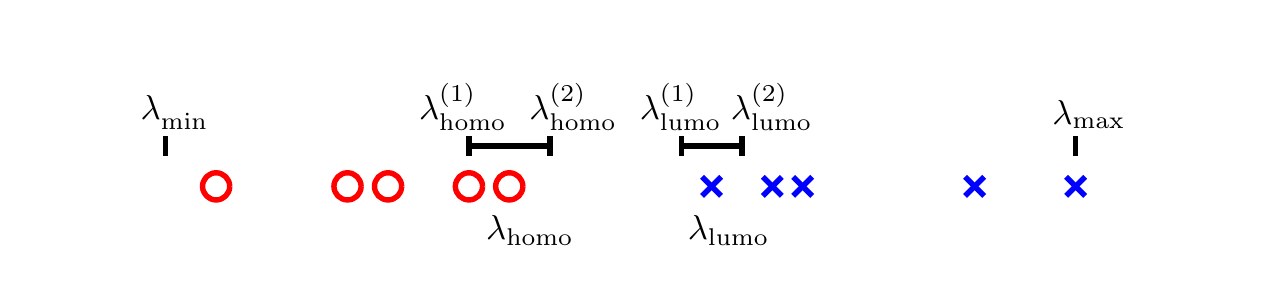}
  \end{center}
  \caption{Eigenvalues of the effective Hamiltonian matrix $H$
    corresponding to occupied (red circles) and unoccupied (blue
    crosses) orbitals.  $(\lambda_{\rm homo}^{(1)}, \ \lambda_{\rm
      homo}^{(2)})$ and $(\lambda_{\rm lumo}^{(1)}, \ \lambda_{\rm
      lumo}^{(2)})$ are intervals containing the homo and lumo
    eigenvalues, respectively. The chemical potential $\mu$ is a
     number between $\lambda_{\rm homo}$ and $\lambda_{\rm
      lumo}$.  $\lambda_{\rm min}$ and $\lambda_{\rm max}$ are either
    the extremal eigenvalues or lower and upper bounds thereof.
    We will refer to $\lambda_{\rm homo}^{(2)}$ and $\lambda_{\rm
      lumo}^{(1)}$ as \emph{inner bounds}, and $\lambda_{\rm
      homo}^{(1)}$ and $\lambda_{\rm lumo}^{(2)}$ as \emph{outer
      bounds}.
    \label{fig:eig_fig}}
\end{figure}

The single-particle density matrix at zero electronic temperature can
be defined as a matrix function of the effective Hamiltonian matrix,
\begin{equation} \label{eq:heavi}
  D = \theta (\mu I - H),
\end{equation}
where $\theta(x)$ is the Heaviside step function and $\mu$ is the
chemical potential or Fermi level. We restrict our applications to
non-metallic materials and assume that $H$ does not have degenerate
eigenvalues at $\mu$.  We further assume that all matrices are Hermitian.
We will refer to the smallest eigenvalue above
$\mu$ as the lowest unoccupied molecular orbital (lumo) eigenvalue and
to the largest eigenvalue below $\mu$ as the highest occupied
molecular orbital (homo) eigenvalue.  The gap between the homo and
lumo eigenvalues is referred to as the homo-lumo gap. See
Figure~\ref{fig:eig_fig} for an illustration.

The density matrix is the matrix for orthogonal projection onto the
occupied subspace spanned by the eigenvectors of $H$ that correspond
to eigenvalues smaller than $\mu$. The dimension of the occupied
subspace is equal to the number of occupied electron orbitals,
$N_\mathrm{occ}$. Usually, the number of electrons is given as input to the
program which has to automatically adjust the chemical potential so
that a correct occupation number is achieved.

In general, the method of choice for computation of matrix functions
of symmetric matrices is diagonalization \cite[p.~84]{book-higham}.
In short, an eigendecomposition $H = V \Lambda V^T$ is computed (with
$\Lambda$ diagonal) and the matrix function can then be computed as
$f(H) = V f(\Lambda) V^T$. This is the standard method to solve
\eqref{eq:heavi} in the context of electronic structure calculations.

Since $D$ is a projection matrix, an eigendecomposition, which
possesses information about all eigenpairs of $H$, is not really
needed to form $D$. Any basis of the occupied subspace could be used
to construct $D$.  Important for the accuracy however is to resolve
the step at the chemical potential.  In other words, while rotations
of the eigenvectors \emph{within} the occupied (or unoccupied)
subspace do not affect the result at all, rotations corresponding to a
leakage between the occupied and unoccupied subspaces do affect the
accuracy, and are more likely to occur between eigenvectors
corresponding to eigenvalues around the chemical potential. One would
therefore expect the problem to become more difficult when the
homo-lumo gap decreases. This is reflected by the condition number for
the problem.

An appropriate condition number for the problem of computing the
density matrix can be defined as
\begin{equation}
  \kappa = \lim_{h\rightarrow 0} \sup_{A:\|A\| = \Delta\epsilon} \frac{\|\theta(\mu I - (H+hA)) - \theta(\mu I - H)\|}{h},
\end{equation}
which can be evaluated to $\kappa = \Delta\epsilon / \xi$, where
$\Delta\epsilon = \lambda_{\rm max} - \lambda_{\rm min}$ is the
spectral width and $\xi$ is the homo-lumo gap of $H$
\cite{EHRubensson12, accPuri}.  In the assessment of density matrix methods, it
is therefore important to consider how the computational cost scales
with the homo-lumo gap. 

\subsection{The self-consistent field procedure} \label{sec:scf}

In electronic structure methods such as Hartree--Fock
\cite{RMcWeeny60,Roothaan} or density functional theory
\cite{hohen,KohnSham65}, the effective single-particle Hamiltonian
matrix depends on the electronic configuration, which is determined by
the density matrix.  If the density matrix used to construct the
Hamiltonian matrix is equal to the density matrix given by
\eqref{eq:heavi}, we say that the solution is self-consistent.
Finding the correct self-consistent electronic ground state therefore
requires an iterative procedure, with repeated computation of
Hamiltonian and density matrices,
\begin{equation}\label{SCF_proc}
  D_0 \rightarrow H_1 \rightarrow D_1 \rightarrow H_2 \rightarrow D_2
  \rightarrow \dots.
\end{equation}
In its simplest form, this self-consistent field procedure is a
fixed-point iteration, where in each iteration the most recent density
or Hamiltonian matrix is used as input to the next step.  Usually,
however, the Hamiltonian matrix is in each step taken as a linear
combination of the most recent and previous Hamiltonian matrices to
reduce the number of iterations needed to reach convergence.  Popular
mixing schemes include simple linear mixing, the direct inversion in
the iterative subspace (DIIS) method \cite{PPulay80}, and Broyden
mixing \cite{Broyden65,DDJohnson88}.

In Born--Oppenheimer molecular dynamics simulations \cite{DMarx00}, the
self-consistent field procedure is employed in each time step, as the
electronic ground state solution is needed in the computation of the
interatomic forces.  In the context of the present work it
does not matter if the sequence of matrices in (\ref{SCF_proc}) is
generated by a self-consistent field optimization or if it is coming
from molecular dynamics time stepping, as long as two successive
Hamiltonians are reasonably similar to each other. The important point
is that there is some procedure generating a sequence of
Hamiltonian matrices $H_1, H_2, \dots$ for which the corresponding
density matrices need to be computed via \eqref{eq:heavi}.

\section{Recursive polynomial expansion of the density matrix}

There are several alternatives to the eigenvalue decomposition for the
calculation of the matrix step function in \eqref{eq:heavi}.  In the
context of linear scaling electronic structure theory some of the
first techniques were based on serial Chebyshev expansions
\cite{SGoedecker99,SGoedecker94,RSilver94,RSilver96}, where the
sparsity of the Hamiltonian can be used efficiently to achieve linear
scaling complexity in the calculation of the Chebyshev polynomials
through their recurrence relation.  The degree of the Chebyshev
polynomial required to reach a certain accuracy is proportional to
$\kappa = \Delta\epsilon/\xi$ \cite{SGoedecker94}.  The computational cost
therefore scales as $\mathcal{O(\kappa)}$ or at best
$\mathcal{O(\sqrt{\kappa})}$ if the matrix polynomial is explicitly
evaluated with the Paterson--Stockmeyer method \cite{foe-lssbbh, polyEval-ps}.

A difficulty with the Chebyshev techniques is that the chemical
potential in \eqref{eq:heavi} needs to be known in advance or has to be adjusted
iteratively in order to achieve the correct occupation number.  Palser
and Manolopoulos proposed a scheme \cite{APalser98} that overcomes
this problem through a recursive expansion based on the third-order
purification polynomial $3X^2-2X^3$ that originally was
designed by McWeeny to adjust the fractional occupation of approximate
density matrices to represent pure ensembles \cite{hist-mcweeny,
  RMcWeeny60}.  In the Palser--Manolopoulos (PM) scheme (as well as in the
original purification method by McWeeny) the density matrix is given through a
rapid recursive expansion of the Heaviside step function where
\begin{equation} \label{eq:recursion}
  \theta(\mu I -H) = \lim_{n \rightarrow \infty} f_n(f_{n-1}(\ldots
  f_0(H) \ldots)).
\end{equation}
By adjusting $X_0 = f_0(H)$ such that $X_0$ has the correct trace,
i.e. the desired occupation, $N_\mathrm{occ}$, the successive McWeeny
polynomials, $f_n(X_{n-1})$, can be modified to preserve the trace. In
this trace conserving (or canonical) ``purification'' scheme the step is automatically
formed at the chemical potential and no prior knowledge of the
chemical potential is required. By using sparse matrix algebra each
polynomial recursion $X_{n} = f_{n}(X_{n-1})$ can be calculated with
linear scaling effort.

A problem with the Palser--Manolopoulos scheme is that it is slow at
low and high occupation, i.\ e.\ when $N_\mathrm{occ}/N$ is close to 0 or 1.
A solution to this problem was offered by the recursive SP2 expansion
scheme by Niklasson \cite{pur-n}. Instead of using third (or higher)
order polynomials that simultaneously project the occupied eigenvalue spectrum
of $H$ towards $1$ and the unoccupied towards $0$, the SP2 scheme uses a
combination of two second order spectral projection polynomials,
\begin{equation}
f_n(X) = \left\{\begin{array}{l}
X^2\\
2X-X^2
\end{array}\right.
\end{equation}
that either projects the eigenvalues towards $1$ for $f_n(X) = 2X-X^2$
or towards $0$ for $f_n(X) = X^2$, see Figure~\ref{fig:polys}.  It is
easy to see that any recursive combination of these two polynomials
converges to a step function with the step formed in the interval
$[0,1]$. Moreover, since the trace of $X^2$ is always smaller than the
trace of $X$, given that all eigenvalues of $X$ are in $[0,1]$, and
vice versa for $2X-X^2$, we can choose between the two polynomials to
iteratively correct the occupation such that the trace of the
converged expansion automatically has the desired value.  The
occupation correcting SP2 algorithm is described in some detail in
Alg.~\ref{alg:puri_tc2}.  After an initial normalization, where
$\lambda_{\rm max/min}$ are upper and lower bounds of the
eigenspectrum of $H$, all eigenvalues of $X$ are in reverse order in
the interval $[0,1]$. Thereafter we apply the spectral projection
polynomials that are chosen to achieve the correct occupation at
convergence.  Prior knowledge of the chemical potential or the
homo-lumo gap is not required.

Linear scaling computational complexity in density matrix methods is
usually achieved by neglecting matrix elements deemed not to
contribute significantly to the overall accuracy
\cite{mixedNormTrunc}.  For recursive polynomial expansions on the
form \eqref{eq:recursion} the removal of matrix elements can be done
in such a way that given a tolerance $\varepsilon$, an accuracy
$\|D-\widetilde{D}\| \leq \varepsilon$ is guaranteed
\cite{mixedNormTrunc, accPuri}.  Here, $D$ and $\widetilde{D}$ are the
exact and approximate density matrices respectively, and $\|\cdot\|$
is a unitary invariant norm.

The polynomial expansion order increases very rapidly in the SP2
recursion. After only 30 matrix-matrix multiplications the expansion
order is over 1 billion, and the number of required multiplications increases
only with the logarithm of the condition number, i.~e. as
$\mathcal{O}(\log{\kappa})$ \cite{pur-n}.  Furthermore, the SP2 method
performs well wherever the chemical potential may be located. In fact,
the SP2 method is at its best at low and high occupation
numbers. Also, the SP2 method is very memory-efficient as it requires
a total of only two symmetric matrices in memory.

\begin{algorithm}
  \caption{Trace-correcting 2nd-order spectral projection expansion (SP2)}
  \label{alg:puri_tc2}
  \begin{algorithmic}[1]
    \State $X = \frac{\lambda_{\mathrm{max}}I-H}{\lambda_{\mathrm{max}} - \lambda_{\mathrm{min}}}$
    \For{$i = 1,2,\dots,n$}
    \If{$|\mathrm{Tr}(X^2) - N_\mathrm{occ}| < |\mathrm{Tr}(2X-X^2) - N_\mathrm{occ}|$}
    \State $X = X^2$
    \Else
    \State $X = 2X - X^2$
    \EndIf
    \EndFor
    \State \textbf{return} $X$
  \end{algorithmic}
\end{algorithm}

\section{Acceleration} 
The recursive expansion functions in \eqref{eq:recursion} are usually
taken as polynomials that have fixed points at 0 and 1, vanishing
derivatives at 0 and/or 1, and that are monotonically increasing in
the $[0,\, 1]$ interval.  Such polynomials are very good for bringing a
near-idempotent matrix closer to idempotency.  For the SP2
polynomials, for example, the asymptotic convergence rate is
quadratic. However, the initial matrix is typically far from being
idempotent, and most of the work in the recursive expansion is
therefore spent on bringing the matrix near idempotency.  For this
task polynomials on the form described above are not
optimal. By allowing for non-monotonicity in the projection
polynomials the convergence can be boosted \cite{non-monotonic}.

\begin{figure}
  \begin{center}
    \includegraphics[width=0.75\textwidth]{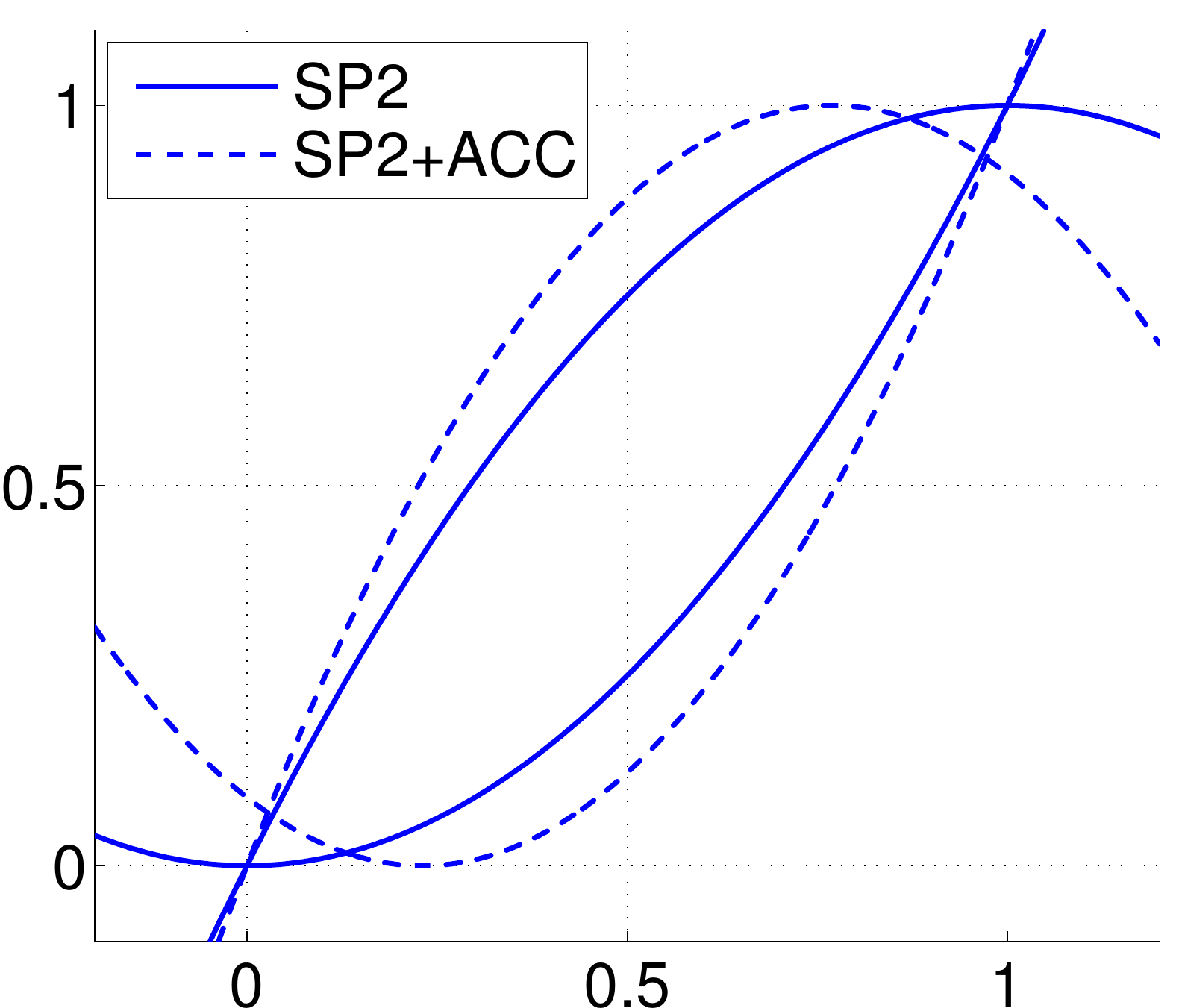}
  \end{center}
  \caption{The regular SP2 projection polynomials $X^2$ and $2X-X^2$
    (solid lines). The shifted and rescaled polynomials used in the
    accelerated recursive expansion, $((1-\alpha)I + \alpha X)^2$ and
    $2\alpha X - (\alpha X)^2$ (dashed lines). Here, the scaling
    parameter was taken as $\alpha = 1.3$. In practice, this parameter
    is adjusted during the course of the recursive expansion to give
    the best possible acceleration, see
    Alg.~\ref{alg:puri_acc_scaled}.
    \label{fig:polys}}
\end{figure}

The basic idea behind the acceleration technique of the recursive
polynomial expansion scheme is that prior knowledge of the eigenvalue
distribution allows for a more optimized design of the projection
polynomials that boosts convergence \cite{non-monotonic}.  What is
needed is, more precisely, estimates of the homo and lumo eigenvalues
$\lambda_{\rm homo}$ and $\lambda_{\rm lumo}$ in
Figure~\ref{fig:eig_fig}. If such estimates are available, the
eigenspectrum can in each step be shifted and scaled so that the
projection polynomials ($X^2$ or $2X-X^2$) fold the eigenspectrum,
which results in a more rapid convergence. Effectively, the SP2 projection
polynomials are shifted and rescaled as illustrated in
Figure~\ref{fig:polys}. The accelerated trace-correcting expansion is
given by Alg.~\ref{alg:puri_acc_scaled}.

Too aggressive scaling can lead to the occupied and unoccupied
eigenstates being mixed up. Therefore, a lower bound of the homo
eigenvalue and an upper bound of the lumo eigenvalue, $\lambda_{\rm
  homo}^{(1)}$ and $\lambda_{\rm lumo}^{(2)}$ in
Figure~\ref{fig:eig_fig}, are used in the accelerated algorithm.  This
makes the algorithm robust with respect to the homo-lumo estimates.
If only loose bounds of the homo and lumo eigenvalues are available,
the acceleration can still be used. As the algorithm is formulated,
loose bounds will never lead to reduced efficiency or accuracy
compared to the original SP2 algorithm, but tighter bounds result in a
more efficient acceleration.

\begin{figure}
  \begin{center}
    \includegraphics[width=0.95\textwidth]{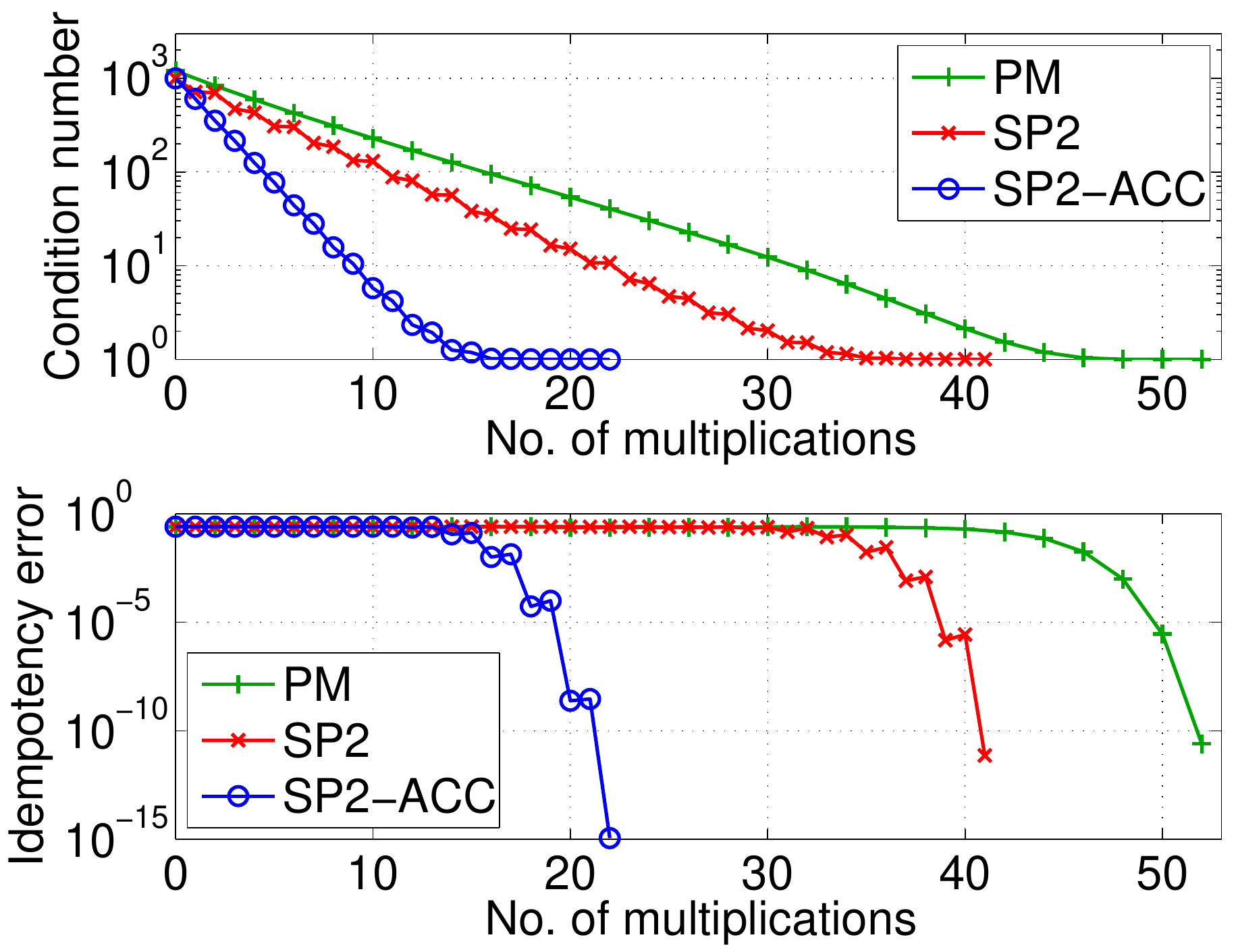}
  \end{center}
  \caption{Illustration of the two phases of the recursive
    expansion. In the conditioning phase, the condition number $\kappa
    = \Delta\epsilon/\xi$ is lowered but the idempotency error
    $\|X-X^2\|_2$ is not reduced. When the condition number is close
    to 1 we have reached the purification phase, where the idempotency
    error decreases quadratically. The test calculations were
    performed for a test Hamiltonian for which the condition number
    $\Delta\epsilon/\xi = 1000$ and with the chemical potential
    located at $\mu = \lambda_{\rm min} + 0.3 (\lambda_{\rm max} -
    \lambda_{\rm min})$. The tested methods are the trace-conserving
    or canonical scheme (PM) \cite{APalser98}, and the
    trace-correcting scheme with (SP2-ACC) and without (SP2)
    acceleration.
    \label{fig:cond_idem}}
\end{figure}

The process of bringing the matrix to idempotency can be seen as
consisting of two different phases, that we shall refer to as the
\emph{conditioning phase} and the \emph{purification phase}. In the
first phase the idempotency error $\|X-X^2\|$ is not reduced, but the
condition number of the problem is lowered.  Note that in each
recursive expansion step, the problem that remains to be solved can be
seen as an independent matrix step function problem with an associated
condition number. The recursive expansion procedure may then be seen
as a procedure to reduce the condition number to 1.  When the
condition number is close to 1, and the conditioning can no longer be
substantially improved, we enter the purification phase in which the
idempotency error is reduced. This behavior is illustrated in
Figure~\ref{fig:cond_idem}.

It is clear that the acceleration represents an improvement of the
conditioning phase which can be explained by a steeper slope of the projection
polynomials at the chemical potential \cite{pur-n}, where the occupied states 
are separated from the unoccupied states.  As the condition number comes close
to 1, and we enter the purification phase, the acceleration parameter
$\alpha_i$ in Alg.~\ref{alg:puri_acc_scaled} comes close to 1 and the method 
becomes essentially equivalent to the original method without acceleration. Thus, the
acceleration is automatically turned off when the purification phase
is reached and the quadratic convergence of the SP2 method sets in.

\begin{algorithm}
  \caption{Accelerated expansion (SP2+ACC)}  \label{alg:puri_acc_scaled}  \begin{algorithmic}[1]
    \State $X = \frac{\lambda_{\mathrm{max}}I-H}{\lambda_{\mathrm{max}} - \lambda_{\mathrm{min}}}$
    \State $x_{1} = \frac{\lambda_{\mathrm{max}}- \lambda_{\rm homo}^{(1)}}{\lambda_{\mathrm{max}} - \lambda_{\mathrm{min}}}$, ~~~~ $~x_{2} = \frac{\lambda_{\mathrm{max}}-\lambda_{\rm lumo}^{(2)}}{\lambda_{ \mathrm{max}} - \lambda_{\mathrm{min}}}$

    \For{$i = 1,2,\dots,n$}
    \If{$|\mathrm{Tr}(X^2) - N_\mathrm{occ}| < |\mathrm{Tr}(2X-X^2) - N_\mathrm{occ}|$}
    \State $p_i = 1$
    \State $\alpha_i = 2 / (2-x_{2})$
    \State $X = (1-\alpha_i)I + \alpha_i X$ \Comment{shift and scale}
    \State $X = X^2$
    \State $\begin{array}{@{}ll}x_{k} =  (1-\alpha_i) + \alpha_i x_{k}, & \\ x_{k} =  x_{k}^2, &  k=1,2 \end{array}$
    \Else
    \State $p_i = 0$
    \State $\alpha_i = 2 / (1+x_{1})$
    \State $X = \alpha_iX$ \Comment{scale}
    \State $X = 2X - X^2$
    \State $\begin{array}{@{}ll} x_{k} = \alpha_i x_{k}, & \\ x_{k} = 2x_{k} - x_{k}^2, & k=1,2\end{array}$
    \EndIf
    \State $v_i = \|X-X^2\|_F$
    \State $w_i = \mathrm{Tr}(X-X^2)$
    \EndFor
    \State \textbf{return} $X$
  \end{algorithmic}

\end{algorithm}

\section{Eigenvalue estimates} 
The drawback with the acceleration technique is that estimates of
interior eigenvalues are needed.  Standard iterative methods, such as
e.g. the Lanczos method, provide rapid convergence to well separated
extremal eigenvalues.  When interior eigenvalues are to be computed,
spectral transformations are typically employed to move the desired
eigenvalues to the ends of the eigenspectrum. Transformations include
shift-and-invert operators and various polynomial filters, see
e.g. \cite{book-eigenvalueTemplates, FangSaad2012}.  The computational
overhead that would be incurred to the recursive expansion by
incorporating iterative eigenvalue methods depends on several factors
such as matrix sparsity and hardware and software details.  In many
cases the computational overhead would be significant.  However, in
the context of recursive polynomial expansions one may take advantage
of the formation of a step at the chemical potential. The recursive
expansion leads to extremely good separation between eigenvalues
around the chemical potential which can be utilized to drastically
reduce the number of Lanczos iterations \cite{intEigs}.  Nevertheless,
even if it is possible to implement such schemes with good efficiency,
the use of \emph{any} iterative method for eigenvalue calculation
would add to the complexity of the method.

Eigenvalue estimates can also be cheaply computed directly from the
entries of the matrix, for example on the basis of Gershgorin's circle
theorem. Unfortunately, Gershgorin's theorem rarely provides useful
information about interior eigenvalues, at least not in the context of the
present work.

In this section, we propose a new simple and efficient approach to calculate
estimates of the homo and lumo eigenvalues when the recursive
expansion is used in an outer loop such as for example molecular
dynamics time stepping.  The proposed method consists of two
parts. The first is calculation of accurate estimates directly from
matrix entries in the recursive expansion of the density matrix. 
The second is propagation of eigenvalue estimates between time steps (or SCF iterations).

\subsection{Accurate interior eigenvalue estimates directly from matrix entries}
The proposed method is based on the following observations. We first
recall that 
\begin{equation} \label{eq:frob_eucl_ineq}
  \frac{\|A\|_F}{\sqrt{N}} \leq \|A\|_2 \leq \|A\|_F.
\end{equation} 
Now, let $\{\lambda_j\}$ be the eigenvalues of $X_i$. Then, 
\begin{equation} \label{eq:frob_norm}
  \|X_i-X_i^2\|_F = \sqrt{\sum_j (\lambda_j - \lambda_j^2)^2}
\end{equation}
and
\begin{equation} \label{eq:eucl_norm}
  \|X_i-X_i^2\|_2 = \max_j{|\lambda_j - \lambda_j^2|}.
\end{equation}

\begin{figure}
  \begin{center}
    \includegraphics[width=0.95\textwidth]{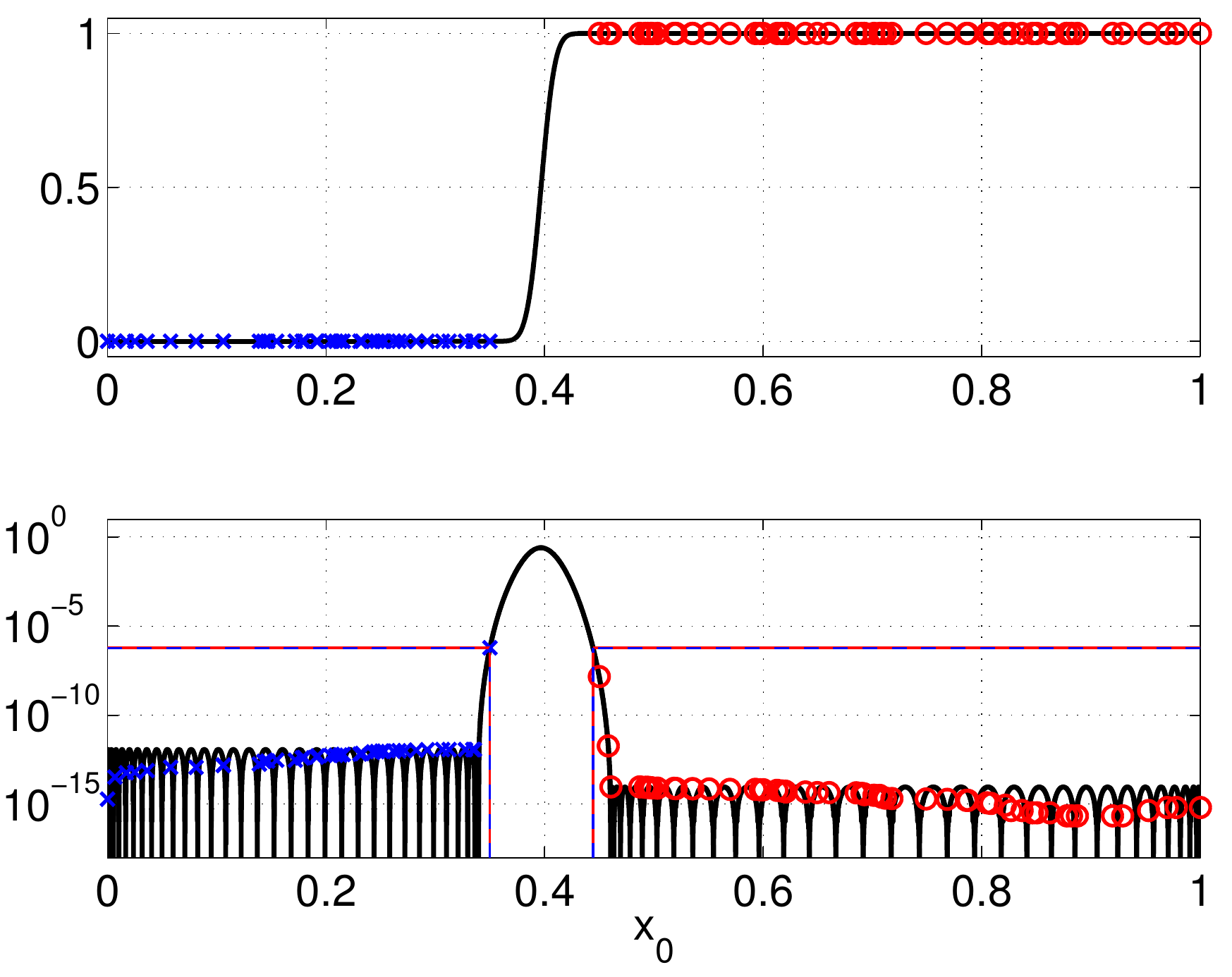}
  \end{center}
  \caption{Mapping of eigenvalues by the accelerated expansion. Upper
    panel: Mapping $f_{12}(f_{11}(\dots f_1(\lambda)\dots))$ of
    occupied (red circles) and unoccupied (blue crosses) eigenvalues
    $\lambda$ of $X_0$.  Lower panel: Absolute values of the
    eigenvalues of $X_{12}-X_{12}^2$ as a function of the eigenvalues
    of $X_0$.  The solid red lines and the dashed blue lines represent
    upper and lower bounds, respectively, of $\|X_{12}-X_{12}^2\|_2$,
    mapped back to the eigenspectrum of $X_0$ to give information
    about the homo and lumo eigenvalues.  The inner bounds for the
    homo and lumo eigenvalues, respectively, are both valid.  In the
    twelfth iteration the outer bound for the lumo eigenvalue also
    holds. The outer bound for the homo eigenvalue, however, is not
    valid.
    \label{fig:illustrate_mapping}}
\end{figure}

The spectral norm in \eqref{eq:eucl_norm} is given by the eigenvalue
closest to 0.5. In the last iterations of the density matrix recursion, 
this eigenvalue is either the homo or the lumo eigenvalue.  Either the homo or the lumo eigenvalue
can then be calculated via calculation of the norm in
\eqref{eq:eucl_norm}.  In the last recursive expansion iterations,
most of the eigenvalues are very close to zero and one, and the sum
over all eigenvalues in \eqref{eq:frob_norm} is dominated by the few
eigenvalues furthest from convergence. This is illustrated in
Figure~\ref{fig:illustrate_mapping}. As we get closer to convergence,
the Frobenius norm therefore becomes a better and better estimate of the
spectral norm.  The Frobenius norm of $X_i-X_i^2$ in the final iterations can 
then be used to obtain estimates to the homo and lumo eigenvalues.  Once the Frobenius norm
has been computed, upper and lower bounds of $\|X_i-X_i^2\|_2$, given
by \eqref{eq:frob_eucl_ineq}, can be translated to bounds for the homo or
lumo eigenvalues by taking the inverse of the recursive expansion \cite{intEigs}.
This gives us a very sharp upper bound. However, for the lower bound in \eqref{eq:frob_eucl_ineq},
we can do better.

Let $\eta_j = \lambda_j - \lambda_j^2, \, j =
1,2,\dots, N$ be the eigenvalues of $X_i-X_i^2$. Since $\eta_j \geq
0,\, j = 1,2,\dots, N$, we have that
\begin{equation}
  \sum_j \eta_j^2 \leq \max_j \eta_j \sum_k \eta_k
\end{equation}
and since 
\begin{eqnarray}
  \sum_j \eta_j^2 & = & \|X_i-X_i^2\|_F^2, \\
  \sum_j \eta_j & = & \textrm{Tr}(X_i-X_i^2)
\end{eqnarray}
and
\begin{equation}
  \|X_i-X_i^2\|_2 = \max_j{\eta_j}
\end{equation}
we have that 
\begin{equation} \label{eq:new_outer}
  \frac{\|X_i-X_i^2\|_F^2}{\textrm{Tr}(X_i-X_i^2)} \leq \|X_i-X_i^2\|_2. 
\end{equation}
Both the lower bound in \eqref{eq:new_outer} and the lower bound 
($\frac{\|X_i-X_i^2\|_F}{\sqrt{N}} \leq \|X_i-X_i^2\|_2$) in \eqref{eq:frob_eucl_ineq}
are sharp in the sense that if all eigenvalues of $X_i-X_i^2$ are equal, then the left hand side equals
the right hand side.  However, the bound in \eqref{eq:new_outer} is
sharp also in the sense that if $X_i-X_i^2$ has only one nonzero
eigenvalue, then the left hand side equals the right hand side as
well. In the present context, many eigenvalues are expected to be
zero, and we therefore expect the bound in \eqref{eq:new_outer} to be much better.

Before we can formulate an algorithm we need to address a few
issues. The first question is which iterations that should be
considered to be the last iterations, so that we can be sure that we
compute the homo or the lumo eigenvalue and not some other eigenvalue.

From Alg.~\ref{alg:puri_acc_scaled}, we have $v_i = \|X_i-X_i^2\|_F$. By
\eqref{eq:frob_eucl_ineq} we have that $v_i \geq \|X_i-X_i^2\|_2 =
\lambda - \lambda^2$ where $\lambda$ is the eigenvalue of $X_i$ that
is closest to 0.5. Assume that $v_i < 1/4$, then we have that the
interval 
\begin{equation} \label{eq:eig_free_interval}
  \left[\frac{1}{2}-\sqrt{\frac{1}{4}-v_i}, \,
    \frac{1}{2}+\sqrt{\frac{1}{4}-v_i}\right]
\end{equation}
 is free from eigenvalues.

Assume now that in iteration $i+1$, $\lambda_\textrm{homo} > 0.5$ and
that $\lambda_\textrm{lumo} < 0.5$, and that the interval $[\gamma, \,
  1-\gamma]$ is free from eigenvalues, where $\gamma$ is a parameter
between 0 and 0.5. If the interval $[\gamma, \, 1-\gamma]$ is free from
eigenvalues also in iteration $i$ and it is not possible for an
eigenvalue to jump from a value larger than $1-\gamma$ to a value
smaller than $\gamma$ and vice versa in one iteration, then we know
that $\lambda_\textrm{homo} > 0.5$ and that $\lambda_\textrm{lumo} <
0.5$ also in iteration $i$.

An eigenvalue can only jump from a value larger than $1-\gamma$ to a
value smaller than $\gamma$ and vice versa in one iteration provided
that $\gamma \geq 6-4\sqrt{2} \approx 0.343146$. This value is given
by solving
\begin{equation}
  \alpha^2\lambda^2 + 2\alpha(1-\alpha)\lambda + (1-\alpha)^2 = \gamma
\end{equation}
where $\lambda = 1-\gamma$ and $\alpha = \frac{2}{2-\gamma}$, for $\gamma \in [0,\, 0.5]$.

Thus, by the above reasoning and specifically by
\eqref{eq:eig_free_interval}, the $v_i$-value can be used for
eigenvalue estimation provided that 
\begin{equation} \label{eq:gamma_criterion}
  v_j < \gamma - \gamma^2, j = i, i+1, \dots, n
\end{equation}
with $\gamma = 6-4\sqrt{2}$.
We only use the $v_i$-values for $i$ that fulfills
this criterion.

It would now be natural to ask the question how to know if it is the
homo or lumo eigenvalue that dominates the Frobenius norm value in
\eqref{eq:frob_norm}.  Our response to this question is that we do not
need to know.  Regardless of whether the homo or lumo eigenvalue
dominates the norm value, $|\lambda_j - \lambda_j^2| \leq v_i$ holds
for all eigenvalues. Since we know from above that
$\lambda_\textrm{lumo} < 0.5$ and that $\lambda_\textrm{homo} > 0.5$,
we therefore have that $\lambda_\textrm{lumo} \leq
\frac{1}{2}(1-\sqrt{1-4v_i})$ and that $\lambda_\textrm{homo} \geq
\frac{1}{2}(1+\sqrt{1-4v_i})$. Thus, the \emph{inner bounds} for the
homo and lumo eigenvalues always hold provided that the
$\gamma$-criterion \eqref{eq:gamma_criterion} is fulfilled.

Clearly, the same is not true for the \emph{outer bounds}, see
Figure~\ref{fig:illustrate_mapping}. However, in each iteration, 
at least one of the outer bounds will be valid, either for the homo or
the lumo eigenvalue.
This means that provided that each of the homo and lumo
eigenvalues dominate the norm value in at least one iteration for
which the $\gamma$-criterion holds, the most conservative outer bound
holds.

When the polynomial in each iteration is chosen based on the trace of
the matrix, as in Algs.~\ref{alg:puri_tc2}
and~\ref{alg:puri_acc_scaled} it can in principle happen that one of
the homo or lumo eigenvalues dominates the norm value in all the final
iterations. This could happen if there are many eigenvalues clustered
very close to either the homo or the lumo eigenvalue.  This problem
can be avoided by choosing polynomial based on the homo and lumo
eigenvalue estimates, as in \cite{accPuri}, which would also improve
convergence. 

Truncation and/or rounding errors cause eigenvalues to be
disturbed. This has the effect that the eigenvalues at convergence
will still deviate from their desired values of $0$ and $1$.  Since
the homo or the lumo eigenvalue dominates the norm value, this only
matters in the very last iterations when the homo and lumo eigenvalues
are very close to $1$ and $0$, respectively. As above this only leads
to too conservative inner bounds and that the outer bounds may not
hold in every iteration, which can be understood from
Figure~\ref{fig:illustrate_mapping}. In our algorithm we use the least
conservative inner bounds and the most conservative outer bounds.

\subsection{Algorithm for homo and lumo eigenvalues}
The algorithm to calculate the estimates is described in
Alg.~\ref{alg:eigs}. 
For each iteration $i$ that the $\gamma$-criterion holds, candidates
for upper and lower bounds of the homo and lumo eigenvalues of $X_i$
are computed.  For those values, the recursive expansion is carried
out backwards so that potential bounds for the homo and lumo
eigenvalues of $X_{i-1}$, $X_{i-2}, \dots$, and eventually $X_0$ are
obtained.  Then, an attempt is made to improve the inner bounds to 
make them tighter and the outer bounds to make them more reliable, see 
the discussion above. Finally, when all potential bounds have been
tested, the selected bounds are translated to bounds for the homo and
lumo eigenvalues of $H$.

The algorithm requires some information that should be collected
during the course of the recursive expansion:
\begin{itemize}
\item[--] The parameters $\lambda_{\mathrm{min}}$ and
  $\lambda_{\mathrm{max}}$ used for the initial normalization.
\item[--] A boolean vector $[p_1,p_2,\dots,p_n]$ specifying which
  polynomial that was used in each iteration.
\item[--] A vector $[\alpha_1, \alpha_2,\dots,\alpha_n]$ containing the
  acceleration or scaling parameters used in each iteration.
\item[--] A vector $[v_1,v_2,\dots,v_n]$ containing the Frobenius norms of
  $X-X^2$ for each iteration.
\item[--] A vector $[w_1,w_2,\dots,w_n]$ containing the traces of 
  $X-X^2$ for each iteration.
\end{itemize}
All this information can be extracted from the recursive expansion
iterations at practically no extra cost, see
Alg.~\ref{alg:puri_acc_scaled}.

\begin{algorithm}
  \caption{Eigenvalue computation}
  \label{alg:eigs}
  \begin{algorithmic}[1]
    \State $i=n$
    \State $x_1 = 1$, ~~~~ $x_2 = 1$, ~~~~ $x_3 = 0$, ~~~~ $x_4 = 0$
    \While{$v_i < \gamma-\gamma^2$}
    \State $y_1 = \frac{1}{2}(1-\sqrt{1-4v_i^2/w_i})$ 
    \State $y_2 = \frac{1}{2}\left(1-\sqrt{1-4v_i}\right)$ 
    \State $y_3 = \frac{1}{2}\left(1+\sqrt{1-4v_i}\right)$ 
    \State $y_4 = \frac{1}{2}(1+\sqrt{1-4v_i^2/w_i})$  
    \For{$j = i,i-1,\dots,2,1$}
    \If{$p_j$}
    \State $\begin{array}{@{}ll} y_k = \sqrt{y_k}, & \\  y_k = (y_k - 1 + \alpha_j)/\alpha_j, & k=1,2,3,4 \end{array}$
    \Else
    \State $\begin{array}{@{}ll} y_k = 1-\sqrt{1-y_k}, & \\ y_k = y_k / \alpha_j, & k=1,2,3,4 \end{array}$
    \EndIf
    \EndFor
    \State $x_k = \min(x_k,y_k), \quad k = 1,2$
    \State $x_k = \max(x_k,y_k), \quad k = 3,4$
    \State $i = i - 1$
    \EndWhile
    \State $\lambda_{\rm homo}^{(1)} = \lambda_{\mathrm{max}} - (\lambda_{\mathrm{max}}-\lambda_{\mathrm{min}})x_4$
    \State $\lambda_{\rm homo}^{(2)} = \lambda_{\mathrm{max}} - (\lambda_{\mathrm{max}}-\lambda_{\mathrm{min}})x_3$
    \State $\lambda_{\rm lumo}^{(1)} = \lambda_{\mathrm{max}} - (\lambda_{\mathrm{max}}-\lambda_{\mathrm{min}})x_2$
    \State $\lambda_{\rm lumo}^{(2)} = \lambda_{\mathrm{max}} - (\lambda_{\mathrm{max}}-\lambda_{\mathrm{min}})x_1$
  \end{algorithmic}
\end{algorithm}

\subsection{Propagation of eigenvalue estimates} \label{sec:propagation}

In the previous section, we developed a method to extract bounds for
the homo and lumo eigenvalues from the recursive polynomial
expansion. This means that the bounds are available only when the
expansion has already completed. In order to accelerate the recursive
expansion the bounds are needed in advance.  However, as
discussed in Section~\ref{sec:scf}, the recursive expansions are often
used repeatedly for a sequence of effective Hamiltonian matrices $H_0,
H_1, \dots$. It is then possible to translate the eigenvalue bounds
from one iteration to the next \cite{accPuri}.

Let $\lambda_1 \leq \lambda_2 \leq \dots \leq \lambda_N$ be the
eigenvalues of $H_{i-1}$ and $\tilde{\lambda}_1 \leq \tilde{\lambda}_2
\leq \dots \leq \tilde{\lambda}_N$ the eigenvalues of $H_{i}$. Then,
by Weyl's theorem on eigenvalue movement \cite{invariant-ss},
\begin{equation} \label{eq:weyl}
  |\lambda_j - \tilde{\lambda}_j| \leq \|H_i - H_{i-1}\|_2, \quad j =
  1,2,\dots, N.
\end{equation}
Therefore, by expanding the intervals containing the homo and lumo
eigenvalues of $H_{i-1}$ with the norm $\|H_i - H_{i-1}\|_2$, we
obtain intervals containing the homo and lumo eigenvalues of $H_{i}$.
If the spectral norm cannot be readily obtained, the Frobenius norm
can be used, see \eqref{eq:frob_eucl_ineq}.  If the norm is small (as
is typically the case between SCF cycles), the acceleration of the
following recursive expansion is hardly affected by this
translation. If the norm is large (as is typically the case between
time steps in a molecular dynamics simulation), the acceleration will
not be optimal, but we have ensured the robustness of the expansion
with respect to the homo and lumo estimates.

\section{Born--Oppenheimer molecular dynamics}
Among computationally
feasible methods for molecular dynamics simulations, Born--Oppenheimer molecular dynamics based on
self-consistent field methods, such as density functional theory
\cite{RMDreizler90,hohen,KohnSham65,RParr89}, is often considered a gold
standard.  In Born--Oppenheimer molecular dynamics the forces acting
on the atoms are explicitly quantum mechanical and calculated at the
relaxed electronic ground state \cite{DMarx00}. The electronic ground
state is given through the self-consistent field optimization procedure, discussed in Section~\ref{sec:scf},
which involves iterative solutions of the density matrix and
Hamiltonian in \eqref{eq:heavi}. 
The self-consistent field optimization accounts for
details in the charge distribution and is important for the accuracy
of the interatomic forces.  

In regular Born--Oppenheimer molecular
dynamics the computational cost, which is dominated by the iterative
self-consistent field optimization, is significantly reduced by using a good initial
guess to the optimization procedure, which is naturally provided through an extrapolation of
the ground state solutions from previous time steps. Since
the iterative self-consistent field optimization in practice never is complete and
always approximate, this extrapolation technique generates a
time-dependent fictitious dynamics of the underlying electronic
degrees of freedom.  Because of the non-linearity of the self-consistent field
optimization this process is irreversible, which typically is
manifested in an unphysical systematic drift in the energy and phase
space \cite{ANiklasson06,PPulay04}, where the electronic degrees of freedom
act like a heat sink or source, constantly removing or adding energy 
to the system.  Only by increasing the number of
self-consistent field iterations can the energy drift be reduced, though it never fully
disappears.  Recently an extended Lagrangian formulation of
Born--Oppenheimer molecular dynamics was introduced that overcomes
these fundamental problems
by restoring time-reversibility and symplecticity to the dynamics
\cite{MCawkwell12,ANiklasson08,ANiklasson12,ANiklasson09,AOdell09,PSteneteg10,GZheng11}.

Instead of using an extrapolation of the electronic ground state 
from previous time steps, as in regular Born--Oppenheimer molecular dynamics, 
an auxiliary electron density $n({\bf r})$ is introduced in the extended Lagrangian
formulation as an extended dynamical variable that is centered around the
ground state density $\rho({\bf r})$ through a harmonic well.  The equations of motion for the
nuclear coordinates ${\bf R} = \{ R_I\}$ and the extended density are given by
\begin{equation}\label{XLBOMD_1}
M_I{\ddot R}_I = -\frac{\partial U({\bf R};\rho)}{\partial R_I},
\end{equation}
\begin{equation}\label{XLBOMD_2}
{\ddot n}({\bf r}) = \omega^2 \left(\rho({\bf r})-n({\bf r})\right),
\end{equation}
where $M_I$ are the nuclear masses and $\omega$ is a frequency
parameter that determines the curvature of the harmonic well.
$U({\bf R};\rho)$ is the potential energy term 
that describes the electronic and ionic energy at
the self-consistent electronic ground state.  The dots denote time-derivatives.

The extended Lagrangian molecular dynamics, \eqref{XLBOMD_1} and \eqref{XLBOMD_2}, has several important
features: {\em i)} $n({\bf r})$ moves in a harmonic well centered around the
ground state density $\rho({\bf r})$ and is therefore a good initial
guess to the iterative self-consistent field optimization procedure, {\em ii)} since $n({\bf
  r})$ is a dynamical variable, in contrast to $\rho({\bf r})$, it can be integrated with a geometric
reversible scheme in the same way as the nuclear coordinates
\cite{BLeimkuhler04,ANiklasson08,AOdell09}, {\em iii)} in this way
reversibility in the underlying propagation of the electronic degrees
of freedom is not broken, and {\em iv)} since the equations of motion for the
nuclear coordinate is identical to ``exact'' Born--Oppenheimer
molecular dynamics, the total Born--Oppenheimer energy,
\begin{equation}\label{Etot}
E_{\rm tot} = \frac{1}{2} \sum_I M_I{\dot R}_I^2 + U({\bf R};\rho),
\end{equation}
is still a constant of motion that should exhibit long-term stability \cite{ANiklasson08}.

Since our estimates of the homo-lumo eigenvalues, which are used for the 
scale-and-fold acceleration in the construction of the density matrix, 
are given from previous time steps, it is important to show that this 
additional artificial propagation does not have any effect on the reversibility
and long-term energy conservation.  Here we will demonstrate how the accelerated SP2
density matrix expansion using estimates of the homo-lumo eigenvalues
from the previous time step is fully compatible with extended
Lagrangian Born--Oppenheimer molecular dynamics, without causing any
systematic drift in the energy.  Our accelerated expansion algorithm
therefore provides a significant reduction of the computational cost,
while keeping the accuracy, for an already highly efficient scheme.

\section{Numerical experiments}

In our numerical experiments the interaction potential $U({\bf R};\rho)$ in \eqref{XLBOMD_1}
is based on the self-consistent density functional
tight-binding (SCC-DFTB) formulation of density functional theory
\cite{MElstner98,MFinnis03,MFinnis98,TFrauenheim00,GZheng05} as
implemented in the program LATTE
\cite{MCawkwell12,MCawkwell12GPU,ANiklasson12,AOdell11,ESanville10}.  
The equations of motion are integrated with the Verlet scheme, in
its symplectic velocity formulation for the nuclear degrees of freedom
and using a regular Verlet scheme for the electronic degrees of freedom
with a dissipative force term that removes any accumulation of
numerical noise \cite{ANiklasson09,PSteneteg10}. 
In all cases the SP2 algorithm, with (Alg.~\ref{alg:puri_acc_scaled}) or without 
(Alg.~\ref{alg:puri_tc2}) acceleration, was used with
the recursions terminated when the occupation change, as measured by the difference
between using $X^2$ or $2X-X^2$, was less than $10^{-10}$.
Two self-consistent field iterations were used in each time step, which together with the
force calculation sums up to totally three density matrix constructions per time step.
The homo and lumo eigenvalue bounds were translated from one time step
or self-consistent field iteration to the next as described in
Section~\ref{sec:propagation}, using the Frobenius matrix norm in \eqref{eq:weyl}. All calculations
were performed using regular dense (non-sparse) matrix algebra.

\begin{figure}
  \begin{center}
    \includegraphics[width=0.95\textwidth]{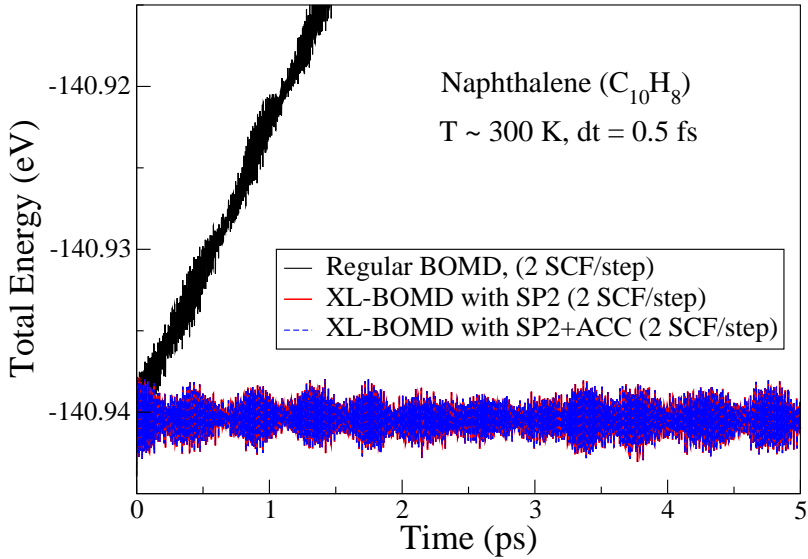}
  \end{center}
  \caption{The conservation of the total energy, \eqref{Etot},
   for regular Born--Oppenheimer molecular dynamics (BOMD) and extended
   Lagrangian Born--Oppenheimer molecular dynamics (XL-BOMD), with and without the
   acceleration scheme (ACC) for the SP2 algorithm. The two lower curves (red and blue)
   are virtually identical.
    \label{fig:etot_cons}}
\end{figure}

Figure \ref{fig:etot_cons} shows the stability of the total
Born--Oppenheimer energy, $E_{\rm tot}$ in \eqref{Etot}, either with regular Born--Oppenheimer molecular 
dynamics using the electronic ground state solution from the previous time step as an
initial guess to the self-consistent field optimization or with extended
Lagrangian Born--Oppenheimer molecular dynamics, \eqref{XLBOMD_1}
and \eqref{XLBOMD_2}, with or without the acceleration scheme.
The systematic drift occurring in regular Born--Oppenheimer
molecular dynamics vanish in the extended Lagrangian formulation.
Moreover, we find virtually no effect in the total energy from using
the accelerated SP2 scheme with estimates of the homo and lumo eigenvalues obtained from the previous iteration. The two graphs are practically 
identical even after thousands of time steps.

\begin{figure}
  \begin{center}
    \includegraphics[width=0.95\textwidth]{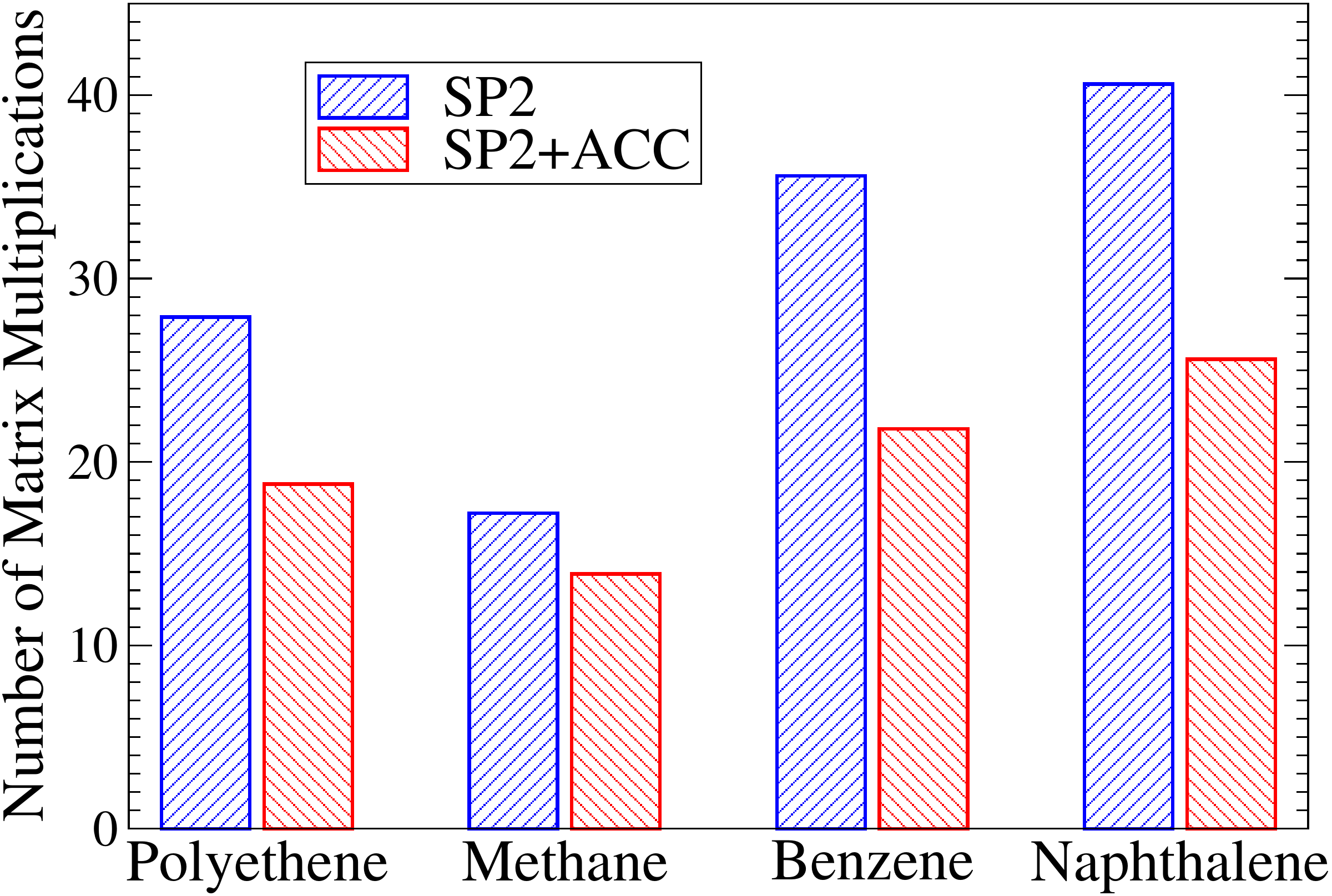} 
  \end{center}
  \caption{The average number of matrix-matrix multiplications of the SP2
   algorithm during molecular dynamics simulations with and without acceleration.
    \label{fig:numb_mm}}
\end{figure}

The gain in efficiency using the acceleration technique is significant.
Figure \ref{fig:numb_mm} shows the average number of matrix-matrix multiplications
required in the density matrix expansion in molecular dynamics simulations of 
various systems. For the simulation of the polyethene molecule we chose a 100 carbon atom
chain (C$_{100}$H$_{202}$) and the methane example consisted of liquid methane (CH$_4$)$_{100}$.
All the simulations were performed with a time step of 0.5 fs using a
structurally optimized ground state configuration with an initial Gaussian 
velocity distribution corresponding to 550 K.
The efficiency gain is governed by the size of the homo-lumo gap \cite{non-monotonic}.
For small gap systems, like naphthalene, the scale-and-fold acceleration gives a significantly higher
reduction of the cost compared to large gap systems, like methane. 
In other words, the acceleration works best for ill-conditioned
systems, as defined in  Section~\ref{sec:density_matrix}.
This also means that there is a much smaller variation in the number
of matrix multiplications required for the accelerated scheme. The
work load is therefore not only smaller but also less sensitive to the
particular choice of materials system. This also means that the computational 
cost is more evenly distributed over time in a molecular dynamics simulation
when there are large variations in the homo-lumo gap.

\begin{figure}
  \begin{center}
    \includegraphics[width=0.95\textwidth]{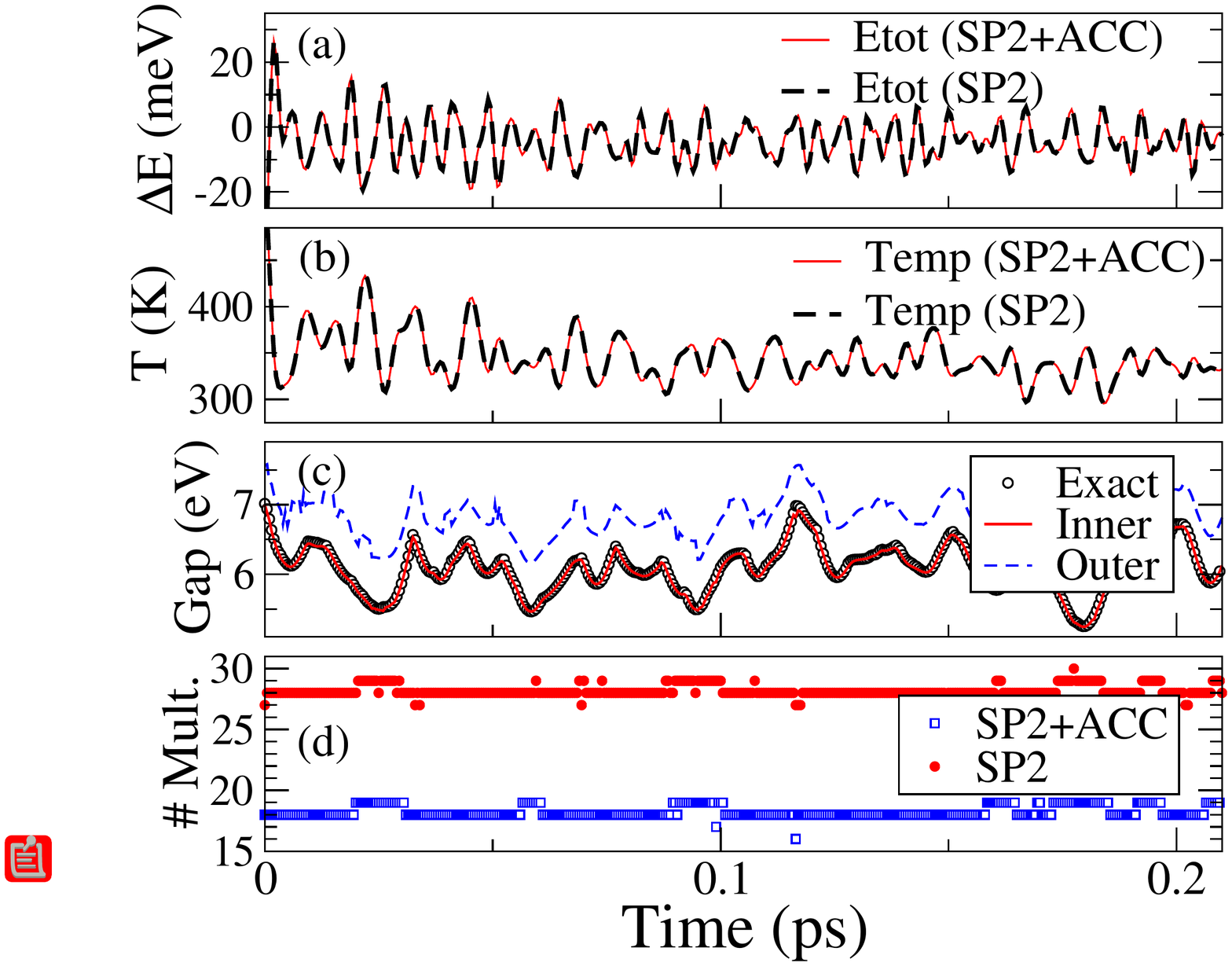}
  \end{center}
  \caption{The fluctuations during a simulation run of liquid isocyanic acid $(HCNO)_{56}$ 
    of the total energy (Etot in panel a), the temperature (Temp in panel b), 
   the gap estimates (panel c), and the variation of the number of matrix-matrix multiplications (panel d)
   of the SP2 algorithm with and without acceleration (ACC).
    \label{fig:isocyn}}
\end{figure}

Figure \ref{fig:isocyn} (panel a-d) illustrates the effect of the
acceleration as a function of simulation time for a system of liquid
isocyanic acid. As seen in the two upper panels (a and b) the
energetics between the regular and the accelerated scheme, as measured
by the total energy or the temperature, are indistinguishable.  Figure
\ref{fig:isocyn} (panel d) also shows how the number of matrix
multiplications for the density matrix construction (that is required
in the last self-consistent field iteration in each time step) varies
over time. Variations in the work load may be sensitive to the choice
of convergence criterion used to terminate the SP2 recursions, which
here is set fairly tight.
The accuracy of the homo-lumo gap estimate, shown in panel (c) is surprisingly good, in
particular as measured by the inner (Inner) $\lambda^{(2)}_{\rm homo}$ and
$\lambda^{(1)}_{\rm lumo}$ eigenvalues.  The upper bound of the gap,
given by the outer (Outer) eigenvalue bounds, is good but not as tight.  
The reason is that for the outer bounds we choose the most conservative
values to make sure that we get valid bounds. 
It is in principle possible to devise a scheme that keeps track of
which eigenvalue (the homo or lumo eigenvalue) that correspond to the
spectral norm of $X-X^2$. This information can be used to obtain more
accurate outer homo-lumo bounds.  However, such an algorithm would be
more complicated with small gains in efficiency.

\section{Discussion}

In this article we have devised a surprisingly simple technique to
extract information about the homo-lumo gap that enables a
cost-efficient practical application of the scale-and-fold
acceleration technique in molecular dynamics simulations. The
technique is quite general and should be applicable to a broad class
of recursive density matrix expansion algorithms besides the SP2
scheme described here \cite{non-monotonic, Suryanarayana2013}.
Scale-and-fold acceleration is not the only way to improve or design
efficient recursive expansion polynomials.  It is in principle
possible to construct a number of ``optimal'' recursive expansion
schemes where the projection polynomials, $\left\{ f_n \right\}$, in
the recursive expansion \eqref{eq:recursion} are defined, for example,
by minimizing the error in some given norm for a certain number of
recursion steps, $N$, and with a maximum number, $M_{\rm max}$, of
necessary matrix multiplications, $M$, e.g.
\begin{equation}\label{OptRec}
{\left\{ f_n \right\}}_{n=0}^{N} = \arg \min_{ {\left\{f_n\right\}} } \left\{ \| \theta(\mu I-H)-f_N(f_{N-1}(\ldots f_0(H) \ldots )) \|;  M \le M_{\rm max} \right\}.
\end{equation}
With a detailed knowledge about the particular eigenvalue spectrum of
$H$ this problem can be solved using a number of various non-linear
optimization techniques.  The key issue with such an optimized
recursion scheme, \eqref{OptRec}, is that we in general do not have
prior knowledge of the interior eigenvalue distribution.  Moreover, in
a molecular dynamics simulation or a self-consistent field optimization
the optimization would have to be performed on-the-fly in each iteration. 
Alternatively, optimizations for different
classes of distributions could be made, e.g. for uniform eigenvalue distributions with 
various homo-lumo gaps and spectral bounds, which possibly could be parametrized. Our accelerated SP2
recursion scheme may very well belong to such a hypothetical class of
optimized expansion methods.

In this article we have taken a cautious standpoint with respect to
the homo and lumo estimates used in the acceleration. Although the
inner bounds of the homo and lumo eigenvalues have proven to be
extremely tight, we have used the more conservative outer bounds for
the acceleration. Moreover, those outer bounds have been moved further
away from the gap as a result of the propagation between time steps or
self-consistent field iterations by~\eqref{eq:weyl}.  All those
measures are not strictly necessary in order to avoid mixing up
occupied and unoccupied states and lead to a less efficient
acceleration.  However, we also have to avoid folding extremal
eigenvalues into the gap since this could lead to invalid homo and
lumo eigenvalue bounds in the following iterations.  With the
conservative measures taken in this work this will never happen,
though in practice it is often possible to use less restrictive
homo-lumo estimates leading to a more efficient acceleration.

\section{Conclusions}
We have proposed a scheme to accelerate the construction of the
density matrix in quantum mechanical molecular dynamics simulations.
Our technique is based on the scale-and-fold acceleration method with
estimates of the humo-lumo gap that are automatically acquired from
the recursive expansion in the previous time step or self-consistent
field iteration at a negligible computational cost.  Our method was
illustrated with density functional tight-binding molecular dynamics
simulations, where the computational effort is dominated by the
density matrix construction.  Our scheme was found to be fully compatible 
with extended Lagrangian Born--Oppenheimer molecular dynamics and 
long-term energy stability.
As a stand-alone technique, our method for calculating the homo-lumo
gap estimates may also be useful in studies of materials properties.

The experiments presented in this work were all carried out using
dense matrix algebra. The use of sparse matrix algebra to achieve a
linear scaling algorithm involves additional issues regarding ways to
bring about matrix sparsity.  Recursive expansions with rigorous error
control requires homo and lumo eigenvalue estimates
\cite{accPuri}. Fortunately, this is the same information that is
needed for the acceleration. Thus, the eigenvalue estimates of the
present work will also be useful in schemes to select small matrix
elements for removal to achieve error control in the linear scaling
construction of the density matrix.

We identified two different phases in recursive density matrix
expansions, the conditioning phase and the purification phase. With
this point of view it was clear that the acceleration technique
represents an improvement of the conditioning phase. In this way, the
acceleration is analogous to preconditioners used with iterative
solvers for linear systems.  By using prior knowledge of the system,
the condition number of the problem is more rapidly reduced in the
(pre-)conditioning phase, which gives a significant reduction of the
computational cost, as demonstrated in quantum mechanical molecular
dynamics simulations.

\section*{Acknowledgments}
Support from G{\"o}ran Gustafsson's foundation, the Swedish research
council (grant no. 621-2012-3861), Lisa and Carl--Gustav Esseens
foundation, the Swedish national strategic e-science research program
(eSSENCE), United States Department of Energy Office of Basic Energy
Sciences, and  the International Ten Bar Java group of T. Peery, as
well as discussions with M. Cawkwell are gratefully acknowledged.

\bibliographystyle{siam}
\bibliography{biblio}

\end{document}